\documentclass[a4paper,11pt]{article}
\pdfoutput=1 

\usepackage{jheppub} 

\usepackage[T1]{fontenc} 

\title{\boldmath Hamiltonian dynamics and gauge symmetry  for three-dimensional Palatini theory with cosmological constant}

\author[a,1]{Alberto Escalante,\note{Corresponding author.}}
\author[b]{Omar Rodr{\'i}guez Tzompantzi}

\affiliation[a]{Instituto de F{\'i}sica, Benem\'erita Universidad Aut\'onoma de Puebla, \\
 Apartado Postal J-48 72570, Puebla Pue., M\'exico,}
\affiliation[b]{Facultad de Ciencias F\'{\i}sico Matem\'{a}ticas, Benem\'erita Universidad Au\-t\'o\-no\-ma de Puebla,
 Apartado postal 1152, 72001 Puebla, Pue., M\'exico.}

\emailAdd{aescalan@ifuap.buap.mx}

\abstract{ A pure Dirac's framework  for 3D  Palatini's   theory with cosmological constant is performed. By considering the complete phase space, we find out  the full structure of the constraints,   and  their corresponding algebra is  computed explicitly.  We report that   in order to obtain a well defined   algebra among the constraints,   the internal group  corresponds to   $SO(2,1)$. In addition, we obtain the extended action,  the extended Hamiltonian,   the   gauge symmetry, and the   Dirac   brackets of the theory. Finally, we compare our results with those reported in the literature. }

\begin{document} 
\maketitle
\flushbottom

\section{INTRODUCTION}
\label{sec:intro}
Models describing  3D  gravity have been used as an alternative  tool  in order to  clarify the highly complex dynamical behavior of the realistic four-dimensional general relativity [GR]. A well-known theory describing  gravity in 3D  is  the so-called  Palatini's theory,  where  the connection and the triad fields are the fundamental dynamical variables. This feature has  been of  great interest in the community for the construction   of a non-perturbative quantum gravity and its  cosmological applications \cite{lec-ashtekar, romano, Lewandowski, Gambini}. In spite that   the Hamiltonian formalism of 3D Palatini's theory has been studied in so many works  with or without  cosmological constant $\Lambda$ \cite{frolov, romano, witten}, there are still difficulties to find out the correct symmetries of the theory, namely, the gauge symmetry   and the  internal   group. In this respect,    some authors have claimed that the gauge symmetry   of 3D [GR]  is Poincar\'e symmetry \cite{witten}, whereas other ones have claimed that it is the Lorentz symmetry plus diffeomorphisms \cite{carlip2}. On the other hand, with respect to the internal group, in  \cite{frolov} it was established that the internal group  is $ISO(2,1)$ Poincar\'e, whereas in \cite{romano} it is reported that 3D Palatini's  theory with or without a cosmological constant $\Lambda$ is well-defined for a wide class of Lie groups; if  $\Lambda=0$, then the internal group $\mathcal{G}$ can be  arbitrary, but  if   $\Lambda\neq0$, then the internal group  $\mathcal{G}$ has to admit an invariant  totally anti-symmetric tensor $\epsilon^{IJK}$. Moreover, if  GR is coupled with matter fields, then the internal group   $\mathcal{G}$  corresponds to $SO(2,1)$. \\
It is well-known that the concept of gauge symmetry plays an important role in modern physics; the physics of the fundamental interactions  based on  the standard model \cite{1},   is a relevant example where the symmetries of a dynamical system  just like gauge covariance is  useful for understanding  the classical and  quantum formulation  of the  theory. In this sense,  the canonical study of  GR is an  important  step to achieve in  order to construct a background independent quantum theory. Thus,  the canonical framework for singular theories is the best tool that we have at hand for studying these  relevant symmetries,  just as  gauge invariance. It is clear  that in any theory presenting  a kind of  symmetries not all of them correspond to  gauge symmetries;  in fact,  the gauge symmetry of any  theory is obtained by following all Dirac's steps \cite{dirac, dirac2, Gitman, Hanson, Henneaux, sundermeyer},  and hence we need to construct,  according to  Dirac's conjecture,  a gauge generator by using the first class constraints. It is worth mentioning  that  usually  Dirac's formalism is not carried out properly, namely, a complete Hamiltonian formulation
means that all steps of the Dirac procedure should be performed, and if  some of these steps are
missing or implemented incorrectly, then we cannot be sure that the correct analysis has been carry out,   and it is possible that the symmetries  found are not correct. In fact, usually the people prefers to work on a smaller phase space  context (also called standard approach); this means that only those variables that occur in the action  with temporal derivative are considered as dynamical,  however, this approach is applicable only when the theory presents certain  simplicity; in general in order to obtain a complete description  one must perform a complete analysis by following all Dirac's steps \cite{alberto, alberto2, alberto3, alberto4, Mu, Rabin}. \\
The aim of the present paper is to perform a pure Dirac's  analysis  for 3D  Palatini's  theory with  cosmological constant. Thus, in our analysis  we shall  consider all the set of one-forms ``$A^{IJ}$'' and ``$e^{I}$'' that define our theory as dynamical ones and not only those variables that occur with  time derivatives in the Lagrangian density,  as  it  is done usually in the standard approach (See Ref. \cite{witten, romano} for a standard analysis). The price to pay by working with a  smaller configuration space, is that we are not able to know the complete form of the constraints, neither the full form of the gauge transformations defined on the complete phase space nor the complete algebra among  the constraints for the theory. Of course,  by working with the full configuration space we can obtain a best and complete description of the theory at a classical level. By working with a complete phase space, we obtain the full structure of the constraints and their full algebra; throughout  this paper, we show that the Poisson algebra among the constraints is closed provided that the internal group  corresponds to $SO(2,1)$. We also show that if we take   the cosmological constant as zero, then the Poisson algebra among  the constraints is closed provided that   the internal group  is still $SO(2,1)$. Hence, it is not necessary to couple matter fields to gravity in order to conclude that  for GR in 3D the internal group   corresponds  to  $SO(2,1)$,  as it is reported in \cite{romano}; in fact, we obtain the same conclusion by performing a detailed canonical analysis without adding  matter fields.  It is important to comment that this result has not been reported in the literature because it is common that some of Dirac's steps are omitted. In this respect,  we can observe in \cite{frolov} a canonical analysis of Palatin's theory without cosmological constant, and  that work  reported that the algebra among  the first class constraints form an   $ISO(2,1) $ algebra, and therefore, the gauge symmetry is Poincar\'e  symmetry;  however, in that work  the second class constraints were solved  and the Dirac brackets were not constructed; we think that those  results are  incomplete. In this paper we obtain the full structure of the first class constraints and the  gauge symmetry for 3D Palatini's theory with or without  cosmological constant; we also  construct the fundamental Dirac's brackets and by using these brackets, we compute  Dirac's algebra among the constraints showing that the algebra is closed. In addition,  we compare our results with those found in the literature. \\
\section{ A pure Dirac's analysis}
It is well known that Palatini's action with  cosmological constant  can be written  as 
\begin{equation}
S[A,e]_{P}=\int_{\mathcal{M}}\epsilon^{IJK}\left[R[A]_{IJ}\wedge e_{K}-\frac{\Lambda}{3}e_{I}\wedge e_{J}\wedge e_{K}\right],
\label{eq:p1}
\end{equation}
where  $A^{IJ}=A_{\mu}{^{IJ}}dx^{\mu}$ is a connection 1-form valued on any Lie group $\mathcal{G}$ that admits an invariant, totally anti-symmetric tensor $\epsilon^{IJK}$ \cite{romano};
$e^{I}=e_{\mu}^{I}dx^{\mu}$ is a triad 1-form that represents the field gravitational, $\Lambda$ is the cosmological constant, and  $R^{IJ}$ is the curvature 2-form of the connection $A^{IJ}$, {\it i.e}.,  $R^{IJ}\equiv dA^{IJ}+A_{\, L}^{I}\wedge A^{LJ}$. Here, $x^{\mu}$ are the coordinates that label the points of the 3-dimensional manifold $\mathcal{M}$. In our notation, Greek letters run from 0 to 2, while  Latin letters  will run from 0 to $\mathfrak{g}$ = dim($\mathcal{G}$). From now on, we will take into account in the number  of dynamical variables,  constraints etc.,  only   the space-time indices, this fact does not affect our results; at the end of our calculations, we will take into account the number of generators of the group.   \\
The equations of motion that arise from the variation of the action (\ref{eq:p1}) with respect to the dynamical  variables are given by
\begin{eqnarray}
\epsilon^{\alpha\mu\nu}\epsilon^{IJK}D_{\mu}e_{\nu{K}} & = & 0,\label{eq:e1}\\
\epsilon^{\alpha\mu\nu}\epsilon_{IJK}\left[R{}_{\mu\nu}{^{JK}}-\lambda e_{\mu}{^{ J}}e_{\nu}{^{ K}}\right]  & = &  0,
\label{eq:p5}
\end{eqnarray}
where (\ref{eq:e1}) is  the zero-torsion condition, which can be solved to get the unique torsion-free spin-connection compatible with $e_{\alpha}{^{I}}$. Inserting the solution of  (\ref{eq:e1}) into (\ref{eq:p5}), one gets Einstein's equation
\begin{equation}
G^{\mu\nu}+\Lambda{g^{\mu\nu}}=0.
\end{equation}
We recall that this equation implies that the space-time has constant scalar curvature proportional
to $\Lambda$, {\it i.e}., $R = 6\Lambda$. Note that this result is independent of the signature of the space-time
and therefore, of the internal group  that we are considering. Moreover, in order to perform the Hamiltonian analysis, we will assume that the manifold $\mathcal{M}$ is topologically $\Sigma\times\mathcal{R}$, where $\Sigma$ corresponds to a Cauchy's surface without boundary $(\partial\Sigma=0)$ and $\mathcal{R}$ represents an evolution parameter.\\
 By performing the $2+1$  decomposition of our fields without  breaking the internal $\mathcal{G}$ symmetry and also without fixing any gauge, we can write the action (\ref{eq:p1}) as
\begin{eqnarray}
S[e,A]_{P} & = & \int dx^{3}\left[\epsilon^{0ab}e_{b}{^{K}}\epsilon_{IJK}\dot{A}_{a}{^{IJ}}-\epsilon^{0ab}e_{b}{^{K}}\epsilon_{IJK}D_{a}A_{0}{^{IJ}}+\frac{1}{2}\epsilon^{0ab}\epsilon_{IJK}e_{0}{^{K}}F_{ab}{^{IJ}}\right.\nonumber \\
 & - & \left.\Lambda\epsilon^{0ab}\epsilon_{IJK}e_{0}{^{I}}e_{a}{^{J}}e_{b}{^{K}}\right],
 \label{eq:p9}
\end{eqnarray}
where $D_{a}A_{b}{^{IJ}}=\partial_{a}A_{b}{^{IJ}}+A_{a}{^{IK}}A_{b}{_{K}}{^{J}}+A_{a}{^{JK}}A_{b}{^{I}}_{K}$
and $F_{ab}{^{IJ}}=\partial_{a}A_{b}{^{IJ}}-\partial_{b}A_{a}{^{IJ}}+A_{a}{^{IK}}A_{b}{_{K}}{^{J}}-A_{b}{^{IK}}A_{a}{_{K}}{^{J}}$.
Here $a,b=1,2$ are space coordinate indices. From (\ref{eq:p9}) we can identify  the following Lagrangian density 
\begin{equation}
{\mathcal{L}}=\epsilon^{0ab}e_{b}{^{K}}\epsilon_{IJK}\dot{A}_{a}{^{IJ}}-\epsilon^{0ab}e_{b}{^{K}}\epsilon_{IJK}D_{a}A_{0}{^{IJ}}+\frac{1}{2}\epsilon^{0ab}\epsilon_{IJK}e_{0}{^{K}}F_{ab}{^{IJ}}-\Lambda\epsilon^{0ab}\epsilon_{IJK}e_{0}{^{I}}e_{a}{^{J}}e_{b}{^{K}}.
\label{eq:p10}
\end{equation}
We have commented above that   usually the Hamiltonian analysis of (\ref{eq:p1}) is carried out  in a reduced phase space \cite{witten, romano}; this means that in those works the   1-forms $A^{IJ}$ and $e^{I}$ that occur in the action with   time derivative are considered as  dynamical variables. However,  in this paper we will   not work in that form, we  shall perform our analysis in concordance with  the background independence of the theory, this means, we shall consider  all  $A^{IJ}$ and $e^{I}$  as our set of dynamical variables which define our theory. Hence, by identifying our set of dynamical variables, a  pure Dirac's method requires to define the momenta $(\Pi^{\alpha}{_{I}}, \Pi^{\alpha\beta}{_{IJ}})$  canonically conjugated to $(e_{\alpha}{^{I}}, A_{\alpha}{^{IJ}})$ \cite{Gitman, Hanson}
 \begin{eqnarray}
\Pi^{\alpha}{_{IJ}}=\frac{\delta{\mathcal{L}}}{\delta\dot{A}_{\alpha}{^{IJ}}},\qquad\Pi^{\alpha}{_{I}}=\frac{\delta{\mathcal{L}}}{\delta\dot{e}_{\alpha}{^{I}}}. 
\label{eq:p11}
\end{eqnarray}
On the other hand, the matrix elements of the Hessian
\[
\frac{\partial^{2}{\mathcal{L}}}{\partial(\partial_{\mu}e_{\alpha}{^{I}})\partial(\partial_{\mu}e_{\beta}{^{I}})},\quad\frac{\partial^{2}{\mathcal{L}}}{\partial(\partial_{\mu}e_{\alpha}^{I})\partial(\partial_{\mu}A_{\beta}{^{IJ}})},\quad\frac{\partial^{2}{\mathcal{L}}}{\partial(\partial_{\mu}A_{\alpha}{^{IJ}})\partial(\partial_{\mu}A_{\beta}{^{IJ}})},
\]
are identically zero and the rank of the matrix Hessian is zero. Thus, we expect  $6$ primary constraints.  From the definition of the momenta (\ref{eq:p11}), we identify the following primary constraints
\begin{eqnarray}
\phi^{0}{_{I}} &: = & \Pi^{0}{_{I}}\approx0,\hphantom{1111111}\phi^{a}{_{I}}: = \Pi^{a}{_{I}}\approx0,\nonumber \\
\phi^{0}{_{IJ}} &: = & \Pi^{0}{_{IJ}}\approx0,\hphantom{11111}\text{\ensuremath{\phi^{a}{_{IJ}}}}: = \Pi^{a}{_{IJ}}-\epsilon^{0ab}\epsilon_{IJK}e_{b}{^{K}}\approx0.
\label{eq:p12}
\end{eqnarray}
We can observe that, if  a smaller phase space is considered, the  $\phi^{0}{_{I}}$ and $\phi^{0}{_{IJ}}$ constraints  would not be  taken into account \cite{witten, romano}. However, the purpose of this paper is to work with the complete phase space and so they are crucial for our study.  The canonical Hamiltonian of  the  theory is given by
\begin{equation}
H_{C} = \int\left[-\frac{1}{2}e_{0}{^{K}}\epsilon^{0ab}\epsilon_{IJK}F{^{IJ}}_{ab}-A_{0}{^{IJ}}D_{a}\Pi^{a}{_{IJ}}+{\Lambda}e_{0}{^{I}}e_{a}{^{J}}\Pi{^{a}}_{IJ}\right]dx^{2}.
 \label{eq:p13}
\end{equation}
In this manner, the primary Hamiltonian will be constructed by adding the primary constraints (\ref{eq:p12}) to (\ref{eq:p13}), namely
\begin{equation}
H_{P}=H_{C}+\int\left[\lambda{_{0}}^{I}\phi^{0}{_{I}}+\lambda{_{a}}^{I}\phi^{a}{_{I}}+\lambda{_{0}}^{IJ}\phi^{0}{_{IJ}}+\lambda{_{a}}^{IJ}\phi^{a}{_{IJ}}\right]dx^{2},
\label{eq:p14}
\end{equation}
here $\lambda^{I}{_{0}}, \lambda^{I}{_{a}}, \lambda^{IJ}{_{0}}, \lambda^{IJ}{_{a}}$ are Lagrange multipliers enforcing the constraints. In this theory, the non-vanishing fundamental Poisson brackets are given by
\begin{eqnarray}
\{e_{\alpha}{^{I}}(x),\Pi^{\beta}{_{J}}(y)\} & = & \delta{^{\beta}}_{\alpha}\delta{^{I}}_{J}\delta^{2}(x-y),\nonumber \\
\{A_{\alpha}{^{IJ}}(x),\Pi^{\beta}{_{KL}}(y)\} & = & \frac{1}{2}\delta{^{\beta}}_{\alpha}\left(\delta^{I}{_{K}}\delta^{J}{_{L}}-\delta^{I}{_{L}}\delta^{J}{_{K}}\right)\delta^{2}(x-y).
\label{eq:p15}
\end{eqnarray}
Now, it is  necessary to identify if  the theory have secondary constraints. For this aim, we observe   that the 
$(6 \times 6)$ matrix whose entries are the Poisson brackets of the primary constraints (\ref{eq:p12}), has  rank $=4$ and $2$  null vectors. Therefore, from the consistency conditions and the null vectors we get the following $2$ secondary constraints
\begin{eqnarray}
\dot{\phi}^{0}{_{I}} & = & \{\phi^{0}{_{I}}(x),{H}_{P}\}\hphantom{1}\approx0\quad\Longrightarrow\quad\psi_{I}\hphantom{1}:=-\frac{1}{2}\epsilon^{0ab}\epsilon_{IKL}F_{ab}{^{KL}}+{\Lambda}e_{a}{^{J}}\Pi{^{a}}_{IJ}\approx 0,\nonumber \\
\dot{\phi}^{0}{_{IJ}} & = & \{\phi^{0}{_{IJ}}(x),{H}_{P}\}\approx0\quad\Longrightarrow\quad\psi_{IJ}:=D_{a}\Pi^{a}{_{IJ}}\approx0.
\label{p16}
\end{eqnarray}
The rank   allows us to get  the following expressions for the Lagrange multipliers
\begin{eqnarray}
\dot{\phi}^{a} {_{IJ}}& = & \{\phi^{a}{_{IJ}}(x),{H}_{P}\}\approx0\quad\Longrightarrow\quad\epsilon^{0ab}\epsilon_{IJK}\lambda_{b}{^{K}}\hphantom{1}=2\epsilon^{0ab}\epsilon_{IJK}D_{b}e_{0}{^{K}}+A_{0}{^{ K}}\Pi^{a}{_{JK}}-A{_{0J}}^{ K}\Pi_{IK}^{\quad a},\nonumber \\
\dot{\phi}^{a}{_{I}} & = & \{\phi^{a}{_{I}}(x),{H}_{P}\}\hphantom{1}\approx0\quad\Longrightarrow\quad\epsilon^{0ab}\epsilon_{IJK}\lambda_{ b}{^{JK}}={\Lambda}\Pi^{a}{_{IJ}}e_{0}{^{J}}.
\label{p17}
\end{eqnarray}
In this theory there are not  third  constraints. At this point, we need to identify  the first- and second-class constraints from the primary and secondary ones. In order to achieve this aim, it is necessary to calculate the $[8 \times 8]$ matrix whose entries are the Poisson brackets constructed out of the primary and secondary constraints. The non-vanishing Poisson brackets are given by 
\begin{eqnarray}
\{\phi{_{I}}^{a}(x),\phi{_{KL}}^{b}(y)\} & = & -\epsilon^{0ab}\epsilon_{IKL}\delta^{2}(x-y),\nonumber \\
\{\phi{_{IJ}}^{a}(x),\psi_{K}(y)\} & = & \epsilon^{0ab}\left[-\epsilon_{KIJ}\partial_{b}+\epsilon_{KIL}A_{bJ}{^{L}}+\epsilon_{KLJ}A_{bI}{^{L}}\right]\delta^{2}(x-y),\nonumber \\
\{\phi{_{IJ}}^{a}(x),\psi_{KL}(y)\} & = & \frac{1}{2}\left[\eta_{LJ}\Pi{_{KI}}^{a}-\eta_{LI}\Pi{_{KJ}}^{a}+\eta_{KJ}\Pi{_{IL}}^{a}-\eta_{KI}\Pi{_{JL}}^{a}\right]\delta^{2}(x-y),\nonumber \\
\{\psi_{I}(x),\psi_{J}(y)\} & = & -{\Lambda}\epsilon^{0ab}\epsilon_{IJH}D_{a}e{^{H}}_{b}\delta^{2}(x-y),\nonumber \\
\{\psi_{I}(x),\psi_{KL}(y)\} & = & \frac{1}{2}\left[\eta_{KI}\psi_{L}-\eta_{LI}\psi_{K}+{\Lambda}\left(e_{Ka}\Pi{_{IL}}^{a}-e_{La}\Pi{_{IK}}^{a}\right)\right]\delta^{2}(x-y),\nonumber \\
\{\psi_{IJ}(x),\psi_{KL}(y)\} & = & \frac{1}{2}\left[\eta_{KI}\psi_{LJ}-\eta_{LI}\psi_{KJ}+\eta_{KJ}\psi_{IL}-\eta_{JL}\psi_{IK}\right]\delta^{2}(x-y)\approx0.
\label{p19}
\end{eqnarray}
This matrix has rank $4$ and $4$ null-vectors. Thus, we expect $4$ second-class constraints and $4$ first-class constraints respectively. From the null vectors we can identify the following complete structure of the  first-class constraints
\begin{eqnarray}
\
\gamma^{0}{_{I}} & := & \Pi^{0}{_{I}}\approx0,\hphantom{11111}\gamma_{I}:=-D_{a}\phi^{a}{_{I}}-\frac{1}{2}\epsilon^{0ab}\epsilon_{IKL}F_{ab}{^{KL}}+\Lambda e_{a}{^{J}}\Pi^{a}{_{IJ}}+\Lambda e{_{a}}^{J}\phi{^{a}}_{IJ}\approx0,\nonumber \\
\gamma^{0}{_{IJ}} & := & \Pi^{0}{_{IJ}}\approx0,\hphantom{111}\gamma_{IJ}:=D_{a}\Pi^{a}{_{IJ}}+\frac{1}{2}\epsilon{^{H}}_{IM}\epsilon{^{MF}}_{J}[\phi^{a}{_{F}}e_{aH}-\phi^{a}{_{H}}e_{aF}]\approx0.
\label{p20}
\end{eqnarray}
We can observe that the $\gamma_{I}$ constraint can be thought  as the dynamical constraint whereas the $\gamma_{IJ}$ constraint can be identified with the Gauss constraint for the  theory  as  occurs in Yang-Mills theory. On the other hand, the rank of the matrix (\ref{p19}) yields the following second-class constraints
\begin{eqnarray}
\phi{^a}_{I}:\chi^{a}{_{I}} & = & \Pi^{a}{_{I}}\approx0,\hphantom{11111}\phi^{a}{_{IJ}}: \chi^{a}{_{IJ}}= \Pi^{a}{_{IJ}}-\epsilon^{0ab}\epsilon_{IJK}e_{b}{^{K}}\approx0.
\label{p21}
\end{eqnarray}
It is important to remark that the complete structure of the constraints $\gamma_{I}$ and $\gamma_{IJ}$ given in (\ref{p20}) are fixed through the null vectors  and they are of   first-class. In this way, the method itself allows us to find by means of  the rank and the null vectors of the matrix (\ref{p19}) all the complete structure of the  first- and second-class constraints \cite{Henneaux, dirac, dirac2, sundermeyer, alberto, alberto2, alberto3, alberto4}. This is the advantage of a pure  Dirac method when it is  applied without missing  any step. It is worth mentioning, that in the complete structure of the first class constraints occur the second class constraints,  and  this full structure has not been reported in the literature. In fact, we are able to observe that our constraints and  the constraints reported in \cite{romano, witten, frolov} are not the same. On one hand, in \cite{romano, witten} they work in a  smaller phase space context. On the other hand, in \cite{frolov} they solved the second class constraints before performing  the contraction of the primary and secondary constraints  with the null vectors, hence,  our constraints are different from those  reported in \cite{frolov}. \\
Now, we will observe the implications  obtained by working with a pure Dirac's analysis;   
 the non zero  algebra among all the constraints (\ref{p20}) and (\ref{p21}) is  given by 
\begin{eqnarray}
\{\gamma_{I}(x),\gamma_{J}(y)\} & = & 2\Lambda\gamma_{JI}\delta^{2}(x\text{-y})\approx0, \\
\{\gamma{_{I}}(x),\gamma{_{KL}}(y)\} & = & \frac{1}{2}\left[\eta_{IK}\gamma_{L}-\eta_{IL}\gamma_{K}\right]\delta^{2}(x\text{-y})\approx0, \\
\{\gamma{_{IJ}}(x),\gamma{_{KL}}(y)\} & = & \frac{1}{2}\left[\eta_{IK}\gamma_{LJ}-\eta_{IL}\gamma_{KJ}+\eta_{JK}\gamma_{IL}-\eta_{JL}\gamma_{IK}\right]\delta^{2}(x\text{-y})\approx0, \label{2.19} \\
\{\gamma_{I}(x),\chi^{a}{_{J}}(y)\} & = & 2\Lambda\chi^{a}{_{IJ}}\delta^{2}(x\text{-y})\approx0, \\
\{\gamma_{I}(x),\chi^{a}{_{KL}}(y)\} & = & \frac{1}{2}\left[\eta_{IL}\chi^{a}{_{K}}-\eta_{IK}\chi{_{L}}^{a}\right]\delta^{2}(x\text{-y})\approx0, \\
\{\gamma{_{IJ}}(x),\chi^{a}{_{K}}(y)\} & = & \frac{1}{2}\left[\eta_{KJ}\chi^{a}{_{I}}-\eta_{KI}\chi^{a}{_{J}}\right]\delta^{2}(x\text{-y})\approx0, \\
\{\gamma{_{IJ}}(x),\chi^{a}{_{KL}}(y)\} & = & \frac{1}{2}\left[\eta_{KJ}\chi^{a}{_{IL}}-\eta_{KI}\chi^{a}{_{JL}}+\eta_{LI}\chi^{a}{_{JK}}-\eta_{LJ}\chi^{a}{_{IK}}\right]\delta^{2}(x\text{-y})\approx0, \label{2.23} \\
\{\chi^{a}{_{I}}(x),\chi^{b}{_{KL}}(y)\} & = & -\epsilon^{0ab}\epsilon_{IKL}\delta^{2}(x-y), 
\label{p22}
\end{eqnarray}
Therefore, we can observe from (\ref{2.19}) and (\ref{2.23}) (see the appendix A for details) that the algebra is closed and has a desired form  provided  that  
\begin{equation}
\epsilon^{IJK}\epsilon_{IMN}=(-1)\left(\delta{^{J}}_{M}\delta{^{K}}_{N}-\delta{^{J}}_{N}\delta{^{K}}_{M}\right).
\label{p22.5}
\end{equation}
This is a property of the structure constants  $\epsilon_{JK}{^{I}}=\epsilon^{IMN}\eta_{MJ}\eta_{NK}$ of the Lie algebra of
$SO(2,1)$. Thus, the constraints (\ref{p20}) and (\ref{p21}) are closed under Poisson brackets, i.e., they form  first- and second-class constraint sets  respectively provided that $\mathcal{G}=SO(2,1)$.\\
Furthermore, we can observe that if we  take   $\Lambda\rightarrow0$, the algebra of constraints (\ref{p20}) form a Poincar\'e algebra as  that reported in \cite{frolov}, but the  internal group  is still  $SO(2,1)$  as can be observed from  the algebra of the  constraints  (\ref{2.19}) and (\ref{2.23}) (see the appendix A).  We appreciate a difference among the analysis performed in \cite{frolov} and the analysis of this  work.  In fact, in that work there are not second class constraints and they do not construct the Dirac's brackets. In this paper, we will preserve the second class constraints until the   end of the calculations. At the end of our analysis, we construct   the Dirac's brackets as is shown  below.  We  can also observe that  if  we use the adjoint representation of the Lie algebra of the  group $\mathcal{G}$, then it is not necessary   to restrict  ourselves to $SO(2,1)$ and the algebra among the constraints do not form a Poincar\'e algebra (see Appendix C),  this a clear difference among our results and those reported in the literature. \\
The correct identification of first- and second-class constraints allows us to carry out the counting of degrees of freedom as follows;  there are  12 canonical variables $(e_{\alpha}{^{I}},\Pi^{\alpha}{_{I}},A_{\alpha}{^{IJ}},\Pi^{\alpha}{_{IJ}})$, 4 independent first-class constraints $(\gamma^{0}{_{I}}, \gamma^{0}{_{IJ}}, \gamma_{I}, \gamma_{IJ})$, and 4 independent second-class constraints  $(\chi^{a}{_{I}}, \chi^{a}{_{IJ}})$ (we have taken into account only the  space-time indices ). Thus, we  conclude that 3D Palatini's theory with cosmological constant  lacks of physical degrees of freedom, i.e., it defines a topological field theory as   expected. On the other hand, in order   to obtain the extended action and the fundamental  Dirac's  brackets, we need to  determinate  the unknown Lagrange multipliers.  For this aim, we introduce  the  matrix  $C_{\alpha\beta}$  whose elements are the Poisson brackets of the second-class constraints  given by
\begin{eqnarray}
C_{\alpha\beta} = \left(\begin{array}{cc}
0 & -\epsilon^{0ab}\epsilon_{IJK}\delta^{2}(x-y)\\
\epsilon^{0ab}\epsilon_{IJK}\delta^{2}(x-y) & 0
\end{array}\right),
\label{p22.6}
\end{eqnarray}
the Dirac's bracket of two functionals $A$, $B$ defined on the phase space,  is expressed by
\[
\{F(x),G(y)\}_{D}\equiv\{F(x),G(y)\}+\int d^{2}zd^{2}w\{F(x),\xi^{\alpha}(z)\}C^{-1}{_{\alpha\beta}}\{\xi^{\alpha}(w),G(y)\},
\]
where $\{F(x),G(y)\}$ is the Poisson bracket  between two
functionals $F,G$, and   $\xi^{\alpha}(z)=\left(\chi{_{I}}^{a},\chi{_{IJ}}^{a}\right)$
are the second-class constraints,  and  $C_{\alpha\beta}^{-1}$ is the inverse of (\ref{p22.6}) that has a trivial form. For simplicity we will restrict the Dirac algebra to  the particular case of ${\mathcal{G}} =SO(2,1)$,  by  using this fact,  Dirac's brackets of the dynamical  variables are given by
\begin{eqnarray}
\{e_{a}{^{I}}(x),\Pi^{b}{_{J}}(y)\}_{D} & = & 0,\nonumber\\
\{e_{a}{^{I}}(x),A_{b}{^{KL}}(y)\}_{D} & = & -\frac{1}{2}{\epsilon_{0ab}\epsilon^{IKL}}\delta^{2}(x-y),\nonumber\\
\{A_{b}{^{KL}}(x),e_{a}{^{I}}(x)\}_{D} & = & \frac{1}{2}{\epsilon_{0ab}\epsilon^{IKL}}\delta^{2}(x-y),\nonumber\\
\{A_{a}{^{IJ}}(x),\Pi^{b}{_{KL}}(y)\}_{D} & = & \frac{1}{2}\left(\delta_{K}{^{I}}\delta_{L}{^{J}}-\delta_{L}{^{I}}\delta_{K}{^{J}}\right)\delta_{a}{^{b}}\delta^{2}(x-y).
\label{p29}
\end{eqnarray}
It is well known
that the Dirac brackets (\ref{p29}) are   essential ingredients in the quantization of the
theory \cite{lec-ashtekar}.  By using the brackets given in (\ref{p29}), we also compute the Dirac brackets among the constraints, we show that  the Dirac algebra is closed (see appendix B). For the case of general internal groups, we will report the closure of Dirac's brackets  in forthcoming works.   Furthermore, the number of second class constraints fixes  the number of indeterminate Lagrange multipliers, thus, by using the fact  that the internal group  for the theory is $SO(2,1)$ and  from  (\ref{p17})  we identify the following Lagrange multipliers, 
\begin{eqnarray}
\lambda_{ a}{^{IJ}} := \frac{\Lambda}{4}\epsilon_{0ab}\epsilon^{IJK}\Pi^{ b}{_{KL}}e_{0}{^{L}},\hphantom{11111} \lambda_{ a}{^{I}}  := D_{a}e_{0}{^{I}}+\frac{1}{2}\epsilon_{0ab}\epsilon^{IJK}A_{0J}{^{L}}\Pi^{b}{_{KL}} .
\label{p23}
\end{eqnarray}
Hence,  we use  the Lagrange multipliers (\ref{p23}), the first-class constraints (\ref{p20}), and the second-class constraints (\ref{p21}) in order to identify the extended action for the theory
\begin{eqnarray}
S_{E} & = & \int d^{3}x\left[ \Pi^{0}{_{IJ}}\dot{A}_{0}{^{IJ}}+\Pi^{a}{_{IJ}}\dot{A}_{a}{^{IJ}}+\Pi^{0}{_{I}}\dot{e}_{0}{^{I}}-\Pi^{a}{_{I}}\dot{e}_{a}{^{I}}-\mathcal{H }-\zeta_{0}{^{I}}\gamma^{0}{_{I}}-\zeta_{0}{^{IJ}}\gamma^{0}{_{IJ}}\right.\nonumber \\
 & - & \left.\zeta^{I}\gamma{_{I}}-\zeta^{IJ}\gamma{_{IJ}}-\overline{\lambda}_{a}{^{I}}\chi^{a}{_{I}}-\overline{\lambda}_{ a}{^{IJ}}\chi^{a}{_{IJ}} \right],
\label{p26}
\end{eqnarray}
where $\mathcal{H }$ is a combination of first-class constraints
\begin{equation}
\mathcal{H}=e_{0}{^{I}}\gamma_{I}-A_{0}{^{ IJ}}\gamma_{IJ}\approx0,
\label{p27}
\end{equation}
and $\zeta_{0}{^{I}},\:\zeta_{0}{^{IJ}},\:\zeta^{I},\:\zeta^{IJ},\overline{\lambda}_{a}{^{I}},\overline{\lambda}_{a}{^{IJ}}$,
are the Lagrange multipliers  that enforce the  first- and second-class constraints.
From the extended action we can identify the extended Hamitonian, which is given by
\begin{equation}
H_{E}=\int d^{2}x\left[\mathcal{H}+\zeta_{0}{^{I}}\gamma^{0}{_{I}}+\zeta_{0}{^{IJ}}\gamma^{0}{_{IJ}}+\zeta^{I}\gamma_{I}+\zeta^{IJ}\gamma_{IJ}\right],
\label{p28}
\end{equation}
thus,   the extended Hamiltonian is a linear combination of first-class constraints as expected. It is important to comment,  that  an extended Hamiltonian with the same structure than (\ref{p27}) is reported in  \cite{frolov}, however, we need take into account that our first class constraints have a complete structure. \\
On the other hand,  it is  well-know that in GR  the dynamical  evolution is governed by the constraints reflecting the general covariance of the theory. Moreover, in order to perform the quantization of the theory, it is not possible to construct the Schr\"{o}dinger equation  because the action of the Hamiltonian on physical states is annihilated. The quantization process is carried out by  the implementation of Dirac's quantization program for gauge systems with general covariance as that realized  in Loop Quantum Gravity \cite{lec-ashtekar},  or as it is showed in \cite{Henneaux};  the first class constraints are promoted to operators $\hat{C}_{i}$ on the kinematical Hilbert space and the physical states are those for which the Dirac conditions $\hat{C}_{i}\cdot\Psi= 0$,  are satisfied. Hence, it is mandatory to perform a pure Dirac's analysis in order to identify the complete structure of the constraints, because constraints are the best guideline to perform the quantization. \\
We finish this paper by finding out the gauge symmetry. In fact, one of the most important symmetries present in singular theories with first class constraints  is the gauge symmetry,  because it can help us to identify physical observables \cite{Hanson}. Thus, we need to know explicitly the fundamental  gauge transformations for the  theory. For this aim, we will apply  Castellani's algorithm \cite{Castellani} to construct  the gauge generator. We define the generator of gauge symmetry as
\begin{equation}
G=\int d^{2}x\left[D_{0}\varepsilon^{I}{_{0}}\gamma^{0}{_{I}}+\varepsilon^{I}\gamma_{I}+D_{0}\kappa_{0}{^{IJ}}\gamma^{0}{_{IJ}}+\kappa^{IJ}\gamma_{IJ}\right],
\label{p30}
\end{equation}
Thus, we find that the gauge symmetry  on the phase space are given by 
\begin{eqnarray}
\delta e_{a}{^{I}}& = &D_{a}\varepsilon^{I}+\kappa^{IJ}e_{aJ},\nonumber \\
\delta e_{0}{^{I}}& = &D_{0}\varepsilon_{0}{^{I}},\nonumber \\
\delta A_{a}{^{IJ}}& = &D_{a}\kappa^{JI}+{\Lambda}\varepsilon^{I}e_{a}{^{J}}-{\Lambda}\varepsilon^{J}e_{a}{^{I}},\nonumber \\
\delta A_{0}{^{IJ}}& = &D_{0}\kappa_{0}{^{IJ}},\nonumber \\
\delta\Pi^{a}{_{I}}& = &2\Lambda(\Pi^{a}{_{IJ}}-\epsilon^{0ab}\epsilon_{IJK}e_{b}{^{K}})\varepsilon^{J}+\kappa_{IJ}\Pi^{aJ},\nonumber \\
\delta\Pi^{0}{_{I}} & = &0,\nonumber \\
\delta\Pi^{a}{_{IJ}}& = &\epsilon^{0ab}\epsilon_{IJM}D_{b}\varepsilon^{M}+\frac{1}{2}\left(\varepsilon_{I}\Pi^{a}{_{ J}}-\varepsilon_{J}\Pi^{a}{_{ I}}\right)+\kappa_{I}{^{N}}\Pi^{a}{_{NJ}}-\kappa_{J}{^{N}}\Pi^{a}{_{NI}},\nonumber \\
\delta\Pi^{0}{_{IJ}} & = & 0.
\label{p32}
\end{eqnarray}
However, they can be written in  covariant form by choosing  the parameter in the following  form; $\varepsilon^{I}=\varepsilon_{0}{^{I}}=\Theta^{I}$,
$\kappa_{0}{^{IJ}}=-\kappa^{IJ}=\Delta^{IJ}$, thus, we get  the following gauge symmetry for this theory
\begin{eqnarray}
e_{\mu}{^{I}} & \longrightarrow & e_{\mu}{^{I}}+D_{\mu}\Theta^{I}+\Delta^{IJ}e_{J\mu}.\nonumber \\
A_{\mu}{^{IJ}} & \longrightarrow & A_{\mu}{^{IJ}}+D_{\mu}\Delta^{IJ}+{\Lambda}\Theta^{I}e_{\mu}{^{J}}-{\Lambda}\Theta^{J}e_{\mu}{^{I}}.
\label{p33}
\end{eqnarray}
It is important to remark that (\ref{p33}) correspond to the  gauge symmetry  of the theory but they do not  correspond to diffeomorphisms,  instead they  are  an   $\Lambda$-deformed   Poincar\'e  transformations (see appendix D). Nevertheless,   we can redefine the gauge parameters as
\begin{eqnarray}
\Theta^{I} =  \frac{1}{2}\xi^{\alpha}e_{\alpha}{^{I}}\hphantom{1111} \textrm{and}\hphantom{1111}\Delta^{IJ}  =  \frac{1}{2}\xi^{\alpha}A_{\alpha}{^{IJ}}.
\label{p34}
\end{eqnarray}
In this manner from the gauge symmetry   we obtain 
\begin{eqnarray}
e_{\mu}{^{I}} & \longrightarrow & e_{\mu}{^{I}}+\mathcal{L}_{\xi}e_{\mu}{^{I}}+\frac{1}{2}\xi^{\alpha}\left[D_{\mu}e_{\alpha}{^{I}}-D_{\alpha}e_{\mu}{^{I}}\right],\nonumber \\
A_{\mu}{^{IJ}} & \longrightarrow & A_{\mu}{^{IJ}}+\mathcal{L}_{\xi}A_{\mu}{^{IJ}}+\xi^{\alpha}\left[R_{\mu\alpha}{^{IJ}}-\frac{\Lambda}{2}\left(e_{\mu}{^{I}}e_{\alpha}{^{J}}-e_{\alpha}{^{I}}e_{\mu}{^{J}}\right)\right],
\label{p35}
\end{eqnarray}
which are  (\textit{on-shell})  diffeomorphisms,  and this symmetry is contained in the  gauge symmetry   (\ref{p33}). We observe in (\ref{p33}) that if $\Lambda=0$ we obtain as gauge symmetry  the  Poincar\'e transformations. Thus, we conclude that for GR without cosmological constant  Poisson algebra is closed if  the internal group  $SO(2,1)$  and the gauge symmetries   are  Poincar\'e transformations, extending  the results reported in \cite{frolov}. Furthermore, for GR with cosmological constant Poisson algebra close if the internal group  is $SO(2,1)$  and  the gauge symmetries   are a $\Lambda$-deformed Poincar\'e transformations. The  Dirac  algebra performed for a general internal group is not yet solved, we are working on this subject, which will be reported in elsewhere.  \\
\section{ Summary and conclusions}
In this paper, we have performed a pure Hamiltonian analysis for Palatini's theory with a cosmological constant. In order to obtain the best description of  this theory, all the steps of Dirac's framework  on full configuration space were  followed. By means of the null vectors  and the rank of the matrix whose elements are the Poisson brackets among  primary and secondary  constraints,  we can identify the complete structure of the first and second class constraints.  With the complete structure of the constraints and their algebra,  we  conclude  that the Poisson algebra of  3D  Palatini theory with a cosmological constant $\Lambda$ is well-defined provided  the  internal group  $\mathcal{G}$  is $SO(2,1)$. Moreover, if  the cosmological constant is taken as zero $\Lambda=0$,  then we  observed that  the algebra among first and second class constraints form a Poincar\'e algebra, however,  the algebra is still well defined  by using the fact that   the structure constants are those for  the group   $SO(2,1)$; therefore, our results extend those reported in \cite{witten, romano, frolov}. In fact, we could observe that it is not necessary to couple GR with  matter fields  in order to conclude that the  Poisson algebra is closed provided that the  internal group  is  $SO(2,1)$, independently if matter fields and/or  cosmological constant are  present or not;  in general our results indicate that the Poisson algebra is closed if the  group  of 3D GR written in the first order formalism corresponds to  $SO(2,1)$. We also compute Dirac's algebra among the constraints, we showed that the algebra is closed.  Furthermore, in \cite{frolov} it  is concluded that the gauge symmetry  of  Palatini's theory without a cosmological constant correspond to Poincar\'e symmetry. In fact,  by taking in our results $\Lambda=0$,   the complete structure of the first class constraints found in this work allowed to construct a gauge generator,  and we conclude  that the gauge symmetries   correspond to Poincar\'e transformations confirming  the results reported in \cite{frolov}.  Finally, our results allowed  us  to construct  the fundamental Dirac's brackets of the theory and determine  the full set of Lagrange multipliers, we then constructed  the  extended action.     Therefore, we conclude  this work by pointing out  that it is mandatory to perform a detailed Dirac's analysis in order to identify  the correct symmetries of the theory under study. \\
 We finish with  some  comments. We are able to observe that in Palatini's theory  it is possible  to take the cosmological constant as zero. In fact, the Dirac brackets (\ref{p29}) do not contain terms with the cosmological constant. This fact shows   a  difference among the so-called exotic action for gravity \cite{witten, 11, jaime} and Palatini's  theory.  In fact, in \cite{jaime} a detailed canonical analysis of exotic gravity  is performed, and the following results are reported; there is not any restriction about the internal group, the  gauge symmetry    is  an $\Lambda$-deformed Poincar\'e symmetry. However, the Dirac brackets among the dynamical variables are non-commutative and   the cosmological constant can not vanish. In fact, the Dirac brackets for exotic action are given by \cite{jaime}
\begin{eqnarray}
\{e{^{I}}_{a}(x),e{^{J}}_{b}(y)\}_{D} &=&\frac{1}{\Lambda}\eta{^{IJ}}\epsilon_{0ab}\delta^{2}(x-y), \nonumber \\ 
\{\Pi{^{a}}_{I}(x),\Pi{^{b}}_{J}(y)\}_{D}   &=&\frac{\Lambda}{4}\eta_{IJ}\epsilon{^{0ab}}\delta^{2}(x-y),\nonumber \\ 
\{A{^{IJ}}_{a}(x),\Pi{^{b}}_{LN}(y)\}_{D} &=&\frac{1}{4}\delta{^{b}}_{a}\left[\delta{^{I}}_{L}\delta{^{J}}_{N}-\delta{^{I}}_{N}\delta{^{J}}_{L} \right]\delta^{2}(x-y),\nonumber \\
\{A{^{IJ}}_{a}(x),A{^{LN}}_{b}(y)\}_{D} &=&\frac{1}{2}\left[\eta{^{IL}}\eta{^{JN}}-\eta{^{IN}}\eta{^{JL}}\right]\epsilon{_{0ab}}\delta^{2}(x-y), \nonumber \\ 
\{\Pi{^{a}}_{IJ}(x),\Pi{^{b}}_{LN}(y)\}_{D} &=&\frac{1}{8}\left[\eta_{{IL}}\eta_{{JN}}-\eta_{{IN}}\eta_{{JL}}\right]\epsilon{^{0ab}}\delta^{2}(x-y). 
\label{p31}
\end{eqnarray}
In this manner, we observe a difference at classical level among exotic action for gravity and Palatini's theory (see \cite{jaime}).  In fact, the Dirac brackets (\ref{p29}) are commutative and are different from  (\ref{p31}); in (\ref{p31}) we can not take $\Lambda=0$.\\
 Finally, we would to comment that   recently results confirming  differences among exotic action and  Palatini's gravity in the context of black holes  have  been published, in fact, we can observe in \cite{prl}  results on   exotic black hole giving    some differences of exotic gravity from the normal gravity action.


\appendix
\section{Algebra among the constraints}
In this appendix we develop  the algebra of the constraints $\{\gamma_{I}(x),\chi^{a}{_{J}}(y)\} $ and $\{\chi^{a}{_{I}}(x),\chi^{b}{_{KL}}(y)\} 
$  given by 
\begin{eqnarray}
\{\gamma_{IJ}(x),\chi^{a}{_{KL}}(y)\} 
& = & \frac{1}{2}\left[\eta_{KJ}\Pi^{a}{_{IL}}-\eta_{KI}\Pi^{a}{_{JL}}+\eta_{LI}\Pi^{a}{_{JK}}-\eta_{LJ}\Pi^{a}{_{IK}}-\epsilon^{0ab}\epsilon^{H}{_{IM}}\eta_{JQ}\epsilon^{FMQ}\epsilon_{KLF}e_{bH}\right.\nonumber \\
&-&\left.\epsilon^{0ab}\epsilon_{MFJ}\eta_{IQ}\epsilon^{HQM}\epsilon_{KLH}e_{b}{^{F}}\right]\delta^{2}(x\text{-y})\nonumber \\
& = & \frac{1}{2}\left[\eta_{KJ}\Pi^{a}{_{IL}}-\eta_{KI}\Pi^{a}{_{JL}}+\eta_{LI}\Pi^{a}{_{JK}}-\eta_{LJ}\Pi^{a}{_{IK}}\right.\nonumber \\
&+&\left.\epsilon^{0ab}\epsilon^{H}{_{IM}}\eta_{JQ}\left(\delta^{M}_{K}\delta^{Q}_{L}-\delta^{M}_{L}\delta^{Q}_{K}\right)e_{bH}+\epsilon^{0ab}\epsilon_{MFJ}\eta_{IQ}\left(\delta^{Q}{_{K}}\delta^{M}_{L}-\delta^{Q}_{L}\delta^{M}_{K}\right)e_{b}{^{F}}\right]\delta^{2}(x\text{-y})\nonumber \\
& = & \frac{1}{2}\left[\eta_{KJ}\left(\Pi^{a}{_{IL}}-\epsilon^{0ab}\epsilon^{H}{_{IL}}e_{bH}\right)-\eta_{KI}\left(\Pi^{a}{_{JL}}-\epsilon^{0ab}\epsilon^{H}{_{JL}}e_{bH}\right)\right.\nonumber\\
&+&\left.\eta_{LI}\left(\Pi^{a}{_{JK}}-\epsilon^{0ab}\epsilon^{H}{_{JK}}e_{bH}\right)-\eta_{LJ}\left(\Pi^{a}{_{IK}}-\epsilon^{0ab}\epsilon^{H}{_{IK}}e_{bH}\right)\right]\delta^{2}(x\text{-y})\nonumber \\
& = & \frac{1}{2}\left[\eta_{KJ}\chi^{a}{_{IL}}-\eta_{KI}\chi^{a}{_{JL}}+\eta_{LI}\chi^{a}{_{JK}}-\eta_{LJ}\chi^{a}{_{IK}}\right]\delta^{2}(x\text{-y})\approx0,
\label{om}
\end{eqnarray}
\begin{eqnarray}
\{\gamma_{IJ}(x)&,&\gamma_{KL}(y)\} 
=  \frac{1}{2}\left[\eta_{KI}\left[D_{a}\Pi^{a}{_{LJ}}+\frac{1}{2}\left(\Pi^{a}{_{L}}e_{aJ}-\Pi^{a}{_{J}}e_{aL}\right)\right]-\eta_{LI}\left[D_{a}\Pi^{a}{_{KJ}}+\frac{1}{2}\left(\Pi^{a}{_{K}}e_{aJ}-\Pi^{a}{_{J}}e_{aK}\right)\right]\right.\nonumber \\
&+&\left.\eta_{KJ}\left[D_{a}\Pi^{a}{_{IL}}+\frac{1}{2}\left(\Pi^{a}{_{I}}e_{aL}-\Pi^{a}{_{L}}e_{aI}\right)\right]-\eta_{JL}\left[D_{a}\Pi^{a}{_{IK}}+\frac{1}{2}\left(\Pi^{a}{_{I}}e_{aK}-\Pi^{a}{_{K}}e_{aI}\right)\right]\right]\delta^{2}(x\text{-y})\nonumber \\
& = &\frac{1}{2} \left[\eta_{KI}\gamma_{JL}-\eta_{LJ}\gamma_{KI}+\eta_{KJ}\gamma_{IL}-\eta_{JL}\gamma_{IK}\right]\delta^{2}(x\text{-y})\approx0.
\label{oma}
\end{eqnarray}
where we have used (\ref{p22.5})  and  the following expression in order to obtain a closed algebra
\begin{eqnarray}
\frac{1}{2}\epsilon^{H}{_{IM}}\epsilon^{MF}{_{J}}\left(\Pi^{a}{_{F}}e_{aH}-\Pi^{a}{_{H}}e_{aF}\right) &=& \frac{1}{2}\epsilon_{HIM}\epsilon^{MFQ}\eta_{QJ}\left(\Pi^{a}{_{F}}e_{a}{^{H}}-\Pi^{aH}e_{aF}\right)\nonumber\\
&=&\frac{1}{2}(-1)\left(\delta^{F}_{H}\delta^{Q}_{I}-\delta^{F}_{I}\delta^{Q}_{H}\right) \eta_{QJ}\left(\Pi^{a}{_{F}}e_{a}{^{H}}-\Pi^{aH}e_{aF}\right)\nonumber \\
&=&\frac{1}{2}\left(\Pi^{a}{_{I}}e_{aJ}-\Pi^{a}{_{J}}e_{aI}\right).
\end{eqnarray}
Finally we compute the Poisson bracket among $ \{\gamma_{I}(x),\gamma_{MN}(y)\} $, 
\begin{eqnarray}
\{\gamma_{I}(x),\gamma_{MN}(y)\} 
& = &\{-D_{a}\chi^{a}{_{I}}, D_{a}\Pi^{a}{_{MN}}+\frac{1}{2}\epsilon^{H}{_{MQ}}\epsilon^{QF}{_{N}}\left(\chi^{a}{_{F}}e_{aH}-\chi^{a}{_{H}}e_{aF}\right)\}\nonumber\\
&+& \{-\frac{1}{2}\epsilon^{0ab}\epsilon_{IKL}F_{ab}{^{KL}}, D_{a}\Pi^{a}{_{MN}}+\frac{1}{2}\epsilon^{H}{_{MQ}}\epsilon^{QF}{_{N}}\left(\chi^{a}{_{F}}e_{aH}-\chi^{a}{_{H}}e_{aF}\right)\}\nonumber\\
&+& \{{\Lambda}e_{a}{^{J}}\Pi^{a}{_{IJ}}+{\Lambda}e{_{a}}^{J}\chi{^{a}}_{IJ},D_{a}\Pi^{a}{_{MN}}+\frac{1}{2}\epsilon^{H}{_{MQ}}\epsilon^{QF}{_{N}}\left(\chi^{a}{_{F}}e_{aH}-\chi^{a}{_{H}}e_{aF}\right)\}, \nonumber\\
\label{omar1}
\end{eqnarray}
first, we calculate the following bracket 
\begin{eqnarray}
\star\{-D_{a}\chi^{a}{_{I}},D_{a}\Pi^{a}{_{MN}}&+&\frac{1}{2}\epsilon^{H}{_{MQ}}\epsilon^{QF}{_{N}}\left(\chi^{a}{_{F}}e_{aH}-\chi^{a}{_{H}}e_{aF}\right)\}\nonumber\\
& = & -\frac{1}{2}\epsilon^{H}{_{MQ}}\epsilon^{QF}{_{N}}\{D_{a}\Pi^{a}{_{I}},\Pi^{a}{_{F}}e_{aH}-\Pi^{a}{_{H}}e_{aF}\}\nonumber\\
& = & -\frac{1}{2}\epsilon^{H}{_{MQ}}\epsilon^{QF}{_{N}}\left[\eta_{IF}D_{a}\Pi^{a}{_{H}}-\eta_{IH}D_{a}\Pi^{a}{_{F}}\right]\delta^{2}(x\text{-y})\nonumber\\
&=&\frac{1}{2}\left[\eta_{NI}D_{a}\chi^{a}{_{M}}-\eta_{MI}D_{a}\chi^{a}{_{N}}\right]\delta^{2}(x\text{-y}),
\label{omar2}
\end{eqnarray}
where we have used (\ref{p22.5}). 
\begin{eqnarray}
\star\{-\frac{1}{2}\epsilon^{0ab}\epsilon_{IKL}F_{ab}{^{KL}}, D_{a}\Pi^{a}{_{MN}}&+&\frac{1}{2}\epsilon^{H}{_{MQ}}\epsilon^{QF}{_{N}}\left(\chi^{a}{_{F}}e_{aH}-\chi^{a}{_{H}}e_{aF}\right)\}\nonumber\\
&=&\{-\frac{1}{2}\epsilon^{0ab}\epsilon_{IKL}F_{ab}{^{KL}}, D_{a}\Pi^{a}{_{MN}}\}\nonumber\\
&=&\frac{1}{2}\left[-\frac{1}{2}\epsilon^{0ab}\eta_{MI}\epsilon_{NEL}F_{ab}{^{EL}}+\frac{1}{2}\epsilon^{0ab}\eta_{NI}\epsilon_{MEL}F_{ab}{^{EL}}\right]\delta^{2}(x\text{-y}).\nonumber\\
\end{eqnarray}
Furthermore, we calculate the following bracket 
\begin{eqnarray}
\star\{{\Lambda}e_{a}{^{J}}\Pi^{a}{_{IJ}}&+&{\Lambda}e{_{a}}^{J}\chi{^{a}}_{IJ},D_{a}\Pi^{a}{_{MN}}+\frac{1}{2}\epsilon^{H}{_{MQ}}\epsilon^{QF}{_{N}}\left(\chi^{a}{_{F}}e_{aH}-\chi^{a}{_{H}}e_{aF}\right)\}\nonumber\\
&=&\frac{\Lambda}{2}\left[\eta_{MI}{e_{a}{^{F}}}\Pi^{a}{_{NF}}-\eta_{NI}{e_{a}{^{E}}}\Pi^{a}{_{ME}}+\eta_{MI}e{_{a}}^{F}\chi{^{a}}_{NF}-\eta_{NI}e{_{a}}^{E}\chi{^{a}}_{ME}\right]\delta^{2}(x\text{-y}),\nonumber\\
\end{eqnarray}
where also  we have used (\ref{p22.5}). Therefore we obtain
\begin{equation}
\{\gamma_{I}(x),\gamma_{MN}(y)\}=\frac{1}{2}\left[\eta_{IM}\gamma_{N}-\eta_{IN}\gamma_{M}\right]\delta^{2}(x\text{-y})\approx0,\nonumber \\
\end{equation}
this shows that  the Poisson   algebra is closed provided  the structure constants correspond to the group $SO(2,1)$.

\section{The Dirac brackets among the constraints}
By using the brackets given in (\ref{p29}), we calculate the Dirac brackets among the first class and second class constraints given by 
\begin{eqnarray}
\{\gamma_{I}(x),\gamma_{J}(y)\}_{D} & = & 2\Lambda\left[\gamma_{JI}+\frac{1}{2}\epsilon_{0ab}\left(\epsilon^{M}{_J}{^{L}}\chi^{a}{_{IM}}\chi^{b}{_{L}}-\epsilon^{M}{_I}{^{L}}\chi^{a}{_{JM}}\chi^{b}{_{L}}\right)\right]\delta^{2}(x\text{-y})\approx0,\nonumber \\
\{\gamma{_{I}}(x),\gamma{_{MN}}(y)\}_{D} & = & \frac{1}{2}\left[\eta_{IM}\gamma_{N}-\eta_{IN}\gamma_{M}+\frac{1}{2}\epsilon_{0ab}\left(\epsilon_{I}{^E}{_{M}}\chi^{a}{_{N}}\chi^{b}{_{E}}-\epsilon_{I}{^E}{_{N}}\chi^{a}{_{M}}\chi^{b}{_{E}}\right)\right.\nonumber \\
&+&\left.2\Lambda\epsilon_{0ab}\left(\epsilon{^{KE}}{_{N}}\chi^{a}{_{IK}}\chi^{b}{_{ME}}-\epsilon{^{KE}}{_{M}}\chi^{a}{_{IK}}\chi^{b}{_{NE}}\right)\right]\delta^{2}(x\text{-y})\approx0,\nonumber \\
\{\gamma{_{IJ}}(x),\gamma{_{MN}}(y)\}_{D} & = & \frac{1}{2}\left[\eta_{IM}\gamma_{NJ}-\eta_{IN}\gamma_{MJ}+\eta_{JM}\gamma_{IN}-\eta_{JN}\gamma_{IM}+\frac{1}{2}\epsilon_{0ab}\epsilon_{I}{^D}{_{M}}\left(\chi^{a}{_{J}}\chi^{b}{_{ND}}+\chi^{a}{_{N}}\chi^{b}{_{JD}}\right)\right.\nonumber \\
&+&\frac{1}{2}\epsilon_{0ab}\epsilon_{I}{^D}{_{N}}\left(\chi^{a}{_{J}}\chi^{b}{_{DM}}+\chi^{a}{_{M}}\chi^{b}{_{DJ}}\right)+\frac{1}{2}\epsilon_{0ab}\epsilon_{J}{^D}{_{M}}\left(\chi^{a}{_{I}}\chi^{b}{_{DN}}+\chi^{a}{_{N}}\chi^{b}{_{DI}}\right)\nonumber \\
&+&\left.\frac{1}{2}\epsilon_{0ab}\epsilon_{J}{^D}{_{N}}\left(\chi^{a}{_{I}}\chi^{b}{_{MD}}+\chi^{a}{_{M}}\chi^{b}{_{ID}}\right)\right]\delta^{2}(x\text{-y})\approx0,\nonumber \\
\{\gamma_{I}(x),\chi^{a}{_{J}}(y)\}_{D} & = &0,\nonumber \\
\{\gamma_{I}(x),\chi^{a}{_{KL}}(y)\}_{D} & = & 0,\nonumber \\
\{\gamma{_{IJ}}(x),\chi^{a}{_{K}}(y)\}_{D} & = & 0,\nonumber \\
\{\gamma{_{IJ}}(x),\chi^{a}{_{KL}}(y)\}_{D} & = & 0,\nonumber \\
\{\chi^{a}{_{I}}(x),\chi^{b}{_{KL}}(y)\}_D & = & 0,
\end{eqnarray}
hence the algebra is closed. We can observe that only   squares of   second class constraints  appear. In fact, the Dirac brackets among first class constraints must be square of second class constraints and linear of  first class constraints \cite{Henneaux}, thus, this calculation  shows that our results has a form desired. In addition,    we have used the equation (\ref{p22.5}) in order to obtain that algebra. It is important to comment that we have showed that  Dirac's procedure work with the internal group $SO(2,1)$ and  the corresponding algebra for other internal general groups is not yet solved, however,  we are working on this subject and the results will be reported  in  forthcoming works.  \\

\section{Comments on   standard Dirac's method} 
We have proved that the   Poisson  and Dirac's algebra is closed,  the relation given by $\epsilon^{IJK}\epsilon_{IMN}=(-1)\left(\delta{^{J}}_{M}\delta{^{K}}_{N}-\delta{^{J}}_{N}\delta{^{K}}_{M}\right)$  is a property of the structure constants $ \epsilon{^{I}}_{JK}= \epsilon^{IMN} \eta_{MJ} \eta_{NK}$ of the Lie algebra of $SO(2,1)$.  However, we can observe in \cite{romano} an  analysis for Palatini theory with $\Lambda \neq 0$ performed by using the adjoint representation of the internal group  $\mathcal{G}$, and the Hamiltonian analysis was developed  on a smaller phase space context. In fact, in \cite{romano} it is proved that Palatini theory  (with or without a cosmological constant $\Lambda$ ) is well-defined for a wide class of Lie groups. If $\Lambda=0$, the Lie group $\mathcal{G}$ can be completely arbitrary; if $\Lambda \neq0$, then  $\mathcal{G}$ has to admit an invariant  totally anti-symmetric tensor $\epsilon^{IJK}$. However, in order to  couple degrees of freedom to 3D gravity, for instance a scalar field,  the  algebra among the  first class constraints is closed for   the internal group   $SO(2,1)$. On the other hand, the goal of our paper is that without work with the adjoint representation of the Lie algebra of $\mathcal{G}$ and by using a pure Dirac's analysis,   we have showed  that the Poisson's algebra of first class constraints is closed  provided that  the internal group  is $SO(2,1)$,   and it is not necessary to couple matter fields to 3D gravity in order to obtain those conclusions.  Therefore, if we work with   the adjoint representation of the Lie algebra of $SO(2,1)$,  then 3D Palatini theory with $\Lambda\neq0$  is well-defined for semi-simple Lie groups,  the algebra of the constraints is closed but  does not form a Poincar\'e algebra. Let us show this point;  by using the adjoint representation of the Lie algebra of $SO(2,1)$,  the action (\ref{eq:p1}) takes the following form 
\begin{equation}
S[e,A]= \int_M R[A]^I\wedge e_I-\frac{ \Lambda}{3}\epsilon^{IJK}e_{I}\wedge e_{J}\wedge e_{K}.
\end{equation}
Hence, by performing the Hamiltonian analysis, we find the following first class constraints 
\begin{eqnarray}
\gamma_{I}&=& D_a\Pi ^a_I, \nonumber \\
  \Gamma^I &=&\epsilon^{0ab}F_{ab}^I  -  \Lambda \epsilon_{0ab} \epsilon^{I J K}\Pi^a_J \Pi^a_K, 
\end{eqnarray}
where $(\Pi ^a_I, \pi_K^a  )$ are the momenta canonically conjugated to $(A_a ^I, e_a ^I )$, and $D_a \lambda^I = \partial_a \lambda^I + \epsilon^{I}{_{JK}}A^J  \lambda^K $. The algebra among the first class constraints is given by 
\begin{eqnarray}
\left[\gamma_I,  \gamma_J  \right]&=& \epsilon_{IJ}{^{K}} \gamma_K, \nonumber \\
\left[\gamma_I,  \Gamma^J  \right]&=& \epsilon_{I}{^{JK}} \Gamma_K, \nonumber \\
\left[\Gamma_I,  \Gamma_J  \right]&=& \Lambda \epsilon {_{IJ}}^ K\Gamma_K.
\label{C3}
\end{eqnarray}
In this manner, the algebra among the constraints is closed, however, in order to obtain that algebra  it is not necessary to restrict ourselves  to  $SO(2,1)$. In fact,  by using the adjoint representation of the Lie group $\mathcal{G}$ the algebra  (\ref{C3})  reproduces   the results reported in \cite{romano}, namely, 3D gravity with a cosmological constant  is well-defined for semi-simple Lie groups and the algebra (\ref{C3}) is closed but  does   not form a Poincar\'e algebra. On the other hand, by working without the adjoint representation as is done in our work,  the Poisson  algebra among the constraints is closed provided that $\mathcal{G}=  SO(2,1)$ and the algebra among the constraints form an $\Lambda$-deformed  Poincar\'e algebra. \\
\section{Poincar\'e transformations} 
The gauge symmetries   obtained in   (\ref{p33}), are related with Poincar\'e transformations. In fact,  let us study  the case when the cosmological constant  vanishes;   by considering the following   Lie-algebra valued one-form 
\begin{equation}
\omega_\mu= e{_{\mu}}^I P_I + \frac{1}{2} A{_{\mu}}^{IK}M_{IK},
\end{equation}
where $P_I $ and $M_{IK}$ are the  Poncar\'e generators. By writing $M^I=\frac{1}{2}\epsilon^{IKL} M_{KL} $,  the generators   obey  the standard commutation relations
\begin{eqnarray}
\left[ M^I, M^K \right] &=& \epsilon^{IKL} M_L, \nonumber \\
\left[M^I, P^K\right] &=& \epsilon^{IKL} P_L, \nonumber \\
\left[P^I, P^K\right]&=& 0, 
\end{eqnarray}
here $I,J,K =0, 1, 2$. \\
By considering the  variation of $ \omega$  under the  gauge symmetry  of kind  $\delta \omega= D\lambda = \partial \lambda + [\omega, \lambda] $ where $\lambda= \lambda^I P_I + \frac{1}{2} \lambda^{IK}M_{IK}$, we obtain the following components \cite{Blagojevic}\\
Translations  
\begin{eqnarray}
e_{\mu}{^{I}} & \longrightarrow & e_{\mu}{^{I}}+D_{\mu}\lambda^{I}.\nonumber \\
A_{\mu}{^{IJ}} & \longrightarrow & A_{\mu}{^{IJ}}.
\end{eqnarray}
Lorentz transformations (rotations)
\begin{eqnarray}
e_{\mu}{^{I}} & \longrightarrow & e_{\mu}{^{I}}+ \lambda^{IJ}e_{J\mu}.\nonumber \\
A_{\mu}{^{IJ}} & \longrightarrow & A_{\mu}{^{IJ}}+D_{\mu}\lambda^{IJ}.
\label{p77}
\end{eqnarray}
We can see that  these transformations are those obtained in (\ref{p33}) with $\Lambda=0$. \\
 It is possible  generalise the above results for $\Lambda \neq0$. In fact, now the generators obey the following algebra \cite{Blagojevic}  
\begin{eqnarray}
\left[ M^I, M^K \right] &=& \epsilon^{IKL} M_L, \nonumber \\
\left[M^I, P^K\right] &=& \epsilon^{IKL} P_L, \nonumber \\
\left[P^I, P^K\right]&=& \Lambda \epsilon^{IKL} M_L. 
\end{eqnarray}
By considering  this algebra,   we obtain  the transformation laws  found in Eq. (\ref{p33}).  However, in our work  the  transformations   (\ref{p33})    were obtained by using a pure  Dirac's method.

\acknowledgments
This work was supported by Sistema Nacional de Investigadores (SNI) M\'exico. The authors want to thank R. Cartas-Fuentevilla for reading the manuscript. \\


\end{document}